\documentclass[aps,prl,reprint,groupedaddress]{revtex4-1}
\usepackage{hyperref}
\usepackage{graphicx}

\begin{document}
\title{Band Structure and Effective Masses of Zn$_{1-x}$Mg$_x$O}
\author{Christian Franz} \author{Marcel Giar} \author{Markus Heinemann} \author{Michael Czerner} \author{Christian Heiliger}
\email{Christian.Heiliger@physik.uni-giessen.de} \affiliation{I. Physikalisches Institut, Justus Liebig University, 35392 Giessen, Germany}
\date{\today}
\begin{abstract}
We analyze the influence of the Mg concentration on several important properties of the band structure of Zn$_{1-x}$Mg$_x$O alloys in wurtzite structure using \textit{ab initio} calculations. For this purpose, the band structure for finite concentrations is defined in terms of the Bloch spectral density, which can be calculated within the coherent potential approximation. We investigate the concentration dependence of the band gap and the crystal-field splitting of the valence bands. The effective electron and hole masses are determined by extending the effective mass model to finite concentrations. We compare our results with experimental results and other calculations.
\end{abstract}
\maketitle

\section{Copyright}
This article has been published in a revised form by Cambridge University Press in 
Materials Research Society (MRS) Proceedings Volume \textbf{1494} (2013) \\
Copyright \copyright Materials Research Society 2012\\
\href{http://dx.doi.org/10.1557/opl.2012.1709}{doi: 10.1557/opl.2012.1709}
\section{Introduction}

Zinc oxide is a promising, sustainable material with many prospective applications, especially in opto-electronics. It is well known that the band gap and other properties can be tuned by adding magnesium. For Mg concentrations up to ca. 30\%, the resulting Zn$_{1-x}$Mg$_x$O alloy has wurtzite structure and a direct band gap~\cite{ohtomo1998}. This can be used in multilayer structures to form e.g.\ light-emitting diodes~\cite{choi2010}. Recently, a two-dimensional electron gas with high charge carrier mobility was created in a ZnMgO-ZnO multilayer structure~\cite{koike2004}. This paves the way to new fields of applications like high-frequency and high-power devices. Tsukazaki \textit{et al.}\ were able to measure the integer~\cite{tsukazaki2007} as well as the fractional quantum Hall effect~\cite{tsukazaki2010} in ZnMgO-ZnO heterostructures. While both require a high degree of control of the material properties, the latter is of particular interest from a fundamental research point of view, since it arises from a strongly correlated state with extraordinary properties.

In order to advance these and other applications reliable numerical tools are of great value. Some of the necessary physical parameters like the valence band effective masses are still unknown for finite concentrations. Thus, recent calculations have to resort to linear interpolation between the pure components, or even use the ZnO value for all concentrations~\cite{furno2010}. The most important part of the ZnO band structure, which is the bottom of the conduction band and the top of the valence bands in the vicinity of the $\Gamma$-point, can be well described within the effective mass approximation. We extend this approach to finite Mg concentrations using a Bloch spectral density~\cite{faulkner1980} defined within the coherent potential approximation (CPA)~\cite{soven1967}. Thereby, we provide the bang gap, the valence band splittings, and the electron and hole effective masses for concentrations up to 30\%.

For pure ZnO the band gap and the valence band splittings are well established from experiments \cite{landoltIII,goano2007}. While the band gap is well described by modern \textit{ab initio} methods, there are still open questions on how to compare the calculated masses to experimental results~\cite{schleife2009}.
Ohtomo \textit{et al.}\ were among the first to grow and investigate Zn$_{1-x}$Mg$_x$O solid solution thin films with Mg concentrations up to 33\%~\cite{ohtomo1998}. Among other things, they investigated the concentration dependence of the band gap. Later studies included the cubic phase at high Mg concentrations~\cite{chen2003} and provided information on the exciton binding energies and valence band splittings~\cite{schmidt2003,teng2000}. The electron effective masses for finite concentrations were obtained indirectly by fitting a model equation to experimental data, and different studies obtain rather different concentration dependencies~\cite{cohen2004,lu2006}.

\textit{Ab initio} investigations include a comprehensive series of contributions on various properties of Zn$_{1-x}$Mg$_x$O and Zn$_{1-x}$Cd$_x$O by Schleife \textit{et al}. They calculated the band structure of ZnO and MgO using a HSE03+G$_0$W$_0$ scheme~\cite{schleife2009}. They describe alloys using a cluster expansion, i.e. a supercell calculation with a subsequent statistical treatment employing various thermodynamic models~\cite{schleife2010,schleife2011}. Maznichenko \textit{et al.}\ used the CPA to investigate the structural phase transitions and the fundamental band gaps of Zn$_{1-x}$Mg$_x$O alloys~\cite{maznichenko2009}.
In this contribution we provide results that are difficult to obtain by experiments or other methods including the concentration dependence of the effective masses and the valence band splittings.

\section{Theory}

The results in this paper are obtained using a density functional theory method. We apply the local density approximation (LDA) for the exchange correlation functional. For the self-consistent density and band structure calculation we use an implementation of the KKR-method~\cite{zabloudil}, which employs one-electron Green's functions expanded in spherical harmonics. The alloys are described using the CPA~\cite{soven1967,zabloudil}, which we recently implemented in our KKR-code. The CPA introduces a self-consistent effective medium, which restores the periodicity of a crystal. This allows us to calculate an averaged electron density in $\vec k$-space called Bloch spectral density~\cite{faulkner1980}, which is closely related to the band structure. The CPA allows an accurate description of alloys at a relatively low computational effort. Popescu and Zunger proposed a different method to define an effective band structure of alloys~\cite{popescu2010}. They used a spectral decomposition to extract an alloy band structure from supercell calculations. This has several advantages but requires the calculation of very large supercells.

We consider ZnO and Zn$_{1-x}$Mg$_x$O in wurtzite structure, which is the equilibrium structure of ZnO at ambient conditions. The calculations are performed for two sets of concentration-dependent lattice parameters: fully relaxed and c-plane growth (i.e. with a fixed a-parameter). The fully relaxed lattice parameters are taken from our earlier publication reference~\cite{heinemann2010}, where we used the ABINIT code to calculate the relaxed lattice structure of Zn$_{1-x}$Mg$_x$O supercells for concentrations up to 31\%. Note that we use the LDA results which are listed in Tab.~\ref{tab:lat}. For the c-plane growth we perform a similar calculation but keep the a-parameter fixed during the relaxation. The results are given in Tab.~\ref{tab:lat}, for the computational details see~\cite{heinemann2010}.

\begin{table}
\caption{Mg concentration ($x$) dependent lattice parameters that are used in the calculations: the hexagonal lattice constant $a$, the axes ratio $c/a$ and the parameter $u$. \label{tab:lat}}
\begin{ruledtabular}
\begin{tabular}{l|l|l}
 & Fully relaxed & c-Plane growth \\ \hline
$a(x)/\mathring{A}$ & $3.2180+0.0354\,x$  &  $3.22$  \\
$c(x)/a(x)$ & $1.6096-0.0473\,x$  &  $1.6068-0.0135\,x$  \\
$u(x)$ & $0.3797+0.0097\,x$  &  $0.3800+0.0040\,x$  \\
\end{tabular}
\end{ruledtabular}
\end{table}

For most applications the part of the band structure close to the band gap is most important. Zn$_{1-x}$Mg$_x$O in wurtzite structure has a direct band gap at the $\Gamma$-point. In this energy range the bands are almost parabolic and can be described by the effective mass approximation
\begin{equation} 
E_n(\vec{k}) \approx E_n(\vec 0) + \frac{\hbar^2 \vec{k}^2}{2 m_n^*} ,
\end{equation}

which approximates the band structure of the n-th band $E_n(\vec k)$ by a parabolic band starting at $E_n(\vec 0)$ with an effective mass $m_n^*$  for small $\vec k$. For the single conduction band with spherical symmetry this is a good approximation. The effective masses of the three valence bands differ significantly for directions of $\vec k$ along the $k_z$-axes ($\parallel$) and for directions in the $k_x$-$k_y$-plane ($\perp$). Hence, for the valence bands it is appropriate to introduce two masses
\begin{equation} 
E_n(\vec{k}) \approx E_n(\vec 0) + \frac{\hbar^2 \vec{k}_{\perp}^2}{2 m_{n \perp}^*} + 
\frac{\hbar^2 \vec{k}_{\parallel}^2}{2 m_{n \parallel}^*}.
\end{equation}

\begin{figure*}
\includegraphics[width=\linewidth]{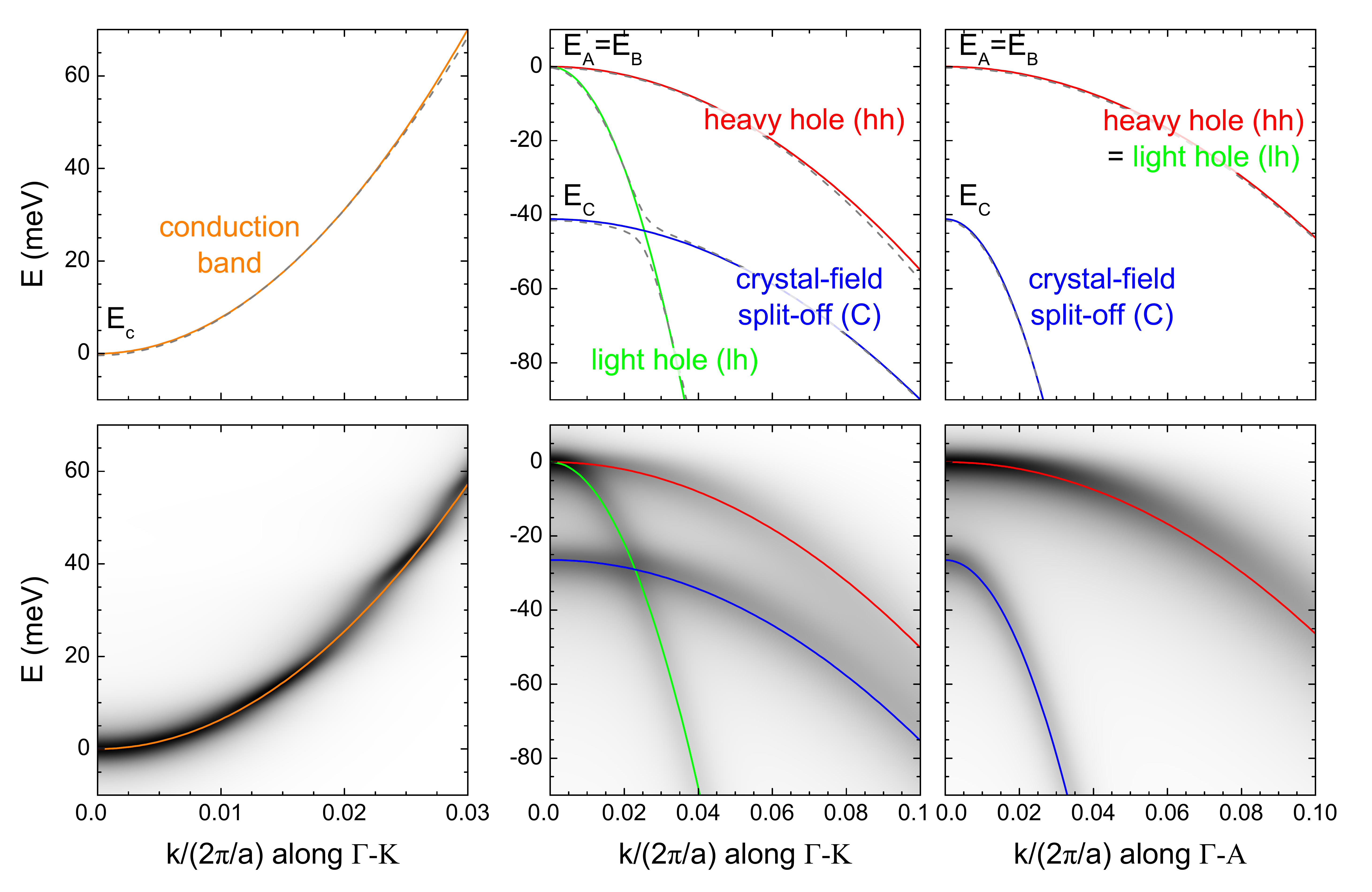}
\caption{\label{img:bands}
Calculated band structure of ZnO (top, dashed line) and the Bloch spectral density of Zn0.85Mg0.15O (bottom, gray gradient), respectively, including the fitted effective mass approximation for the conduction (left) and valence bands (middle, right) for small $\vec{k}$.}
\end{figure*}

The effective mass approximation is illustrated in Fig.~\ref{img:bands}~(top).
The present calculations are carried out without relativistic effects like spin-orbit interaction. The latter introduces a small additional splitting of the valence bands, which results in three distinct bands at the $\Gamma$-point, which are usually labeled $A$, $B$, and $C$. Without spin-orbit interaction the bands $A$ and $B$ are degenerate at the $\Gamma$-point ($E_A(\vec 0)=E_B(\vec 0)$) and we obtain the crystal-field splitting $\Delta_{AC}=E_A(\vec 0)-E_C(\vec 0)$ and the band gap $E_g=E_c(\vec 0)-E_A(\vec 0)$. The $\Gamma$-$A$-line has the same symmetry and degeneracies as the $\Gamma$-point. In the $k_x$-$k_y$-plane the symmetry is lower and the degeneracy between the bands $A$ and $B$ is lifted. In that case these bands have a different dispersion. $A$ has a large effective mass $m_{hh}$ (heavy hole) and $B$ a small effective mass $m_{lh}$ (light hole). The $C$ band is also referred to as crystal-field split-off band ($m_C$). In order to allow for a parabolic fit, the anticrossing of $B$ and $C$ is replaced by a crossing.

For pure ZnO the effective mass approximation can be directly fitted to the calculated band structure (Fig.~\ref{img:bands}, top). For finite concentrations of Mg the disorder leads to a broadening of the bands in the Bloch spectral density shown in Fig.~\ref{img:bands}~(bottom). Despite this broadening, for the concentrations considered here ($x=0.0 - 0.3$), it is possible to fit the effective mass approximation to the density. We consider an energy range of 26 meV for each band for the fitting.

\section{Results and Discussion}

\begin{table}
\caption{Some selected results on various properties of pure ZnO\label{tab:ZnO}}
\begin{ruledtabular}
\begin{tabular}{l|l|l|l}
$E_g\ (\mathrm{eV})$ & $\Delta_{AC}\ (\mathrm{meV})$ & $m_c/m_e$ & $m_{hh\perp}/m_e$ \\
1.08 & 41.2 & 0.186 & 2.64 \\ \hline\hline
$m_{hh\parallel}/m_e$ & $m_{lh\perp}/m_e$ & $m_{C\perp}/m_e$ & $m_{C\parallel}/m_e$ \\
3.13 & 0.212 & 3.03 & 0.208
\end{tabular}
\end{ruledtabular}
\end{table}

\begin{figure}
\includegraphics[width=\linewidth]{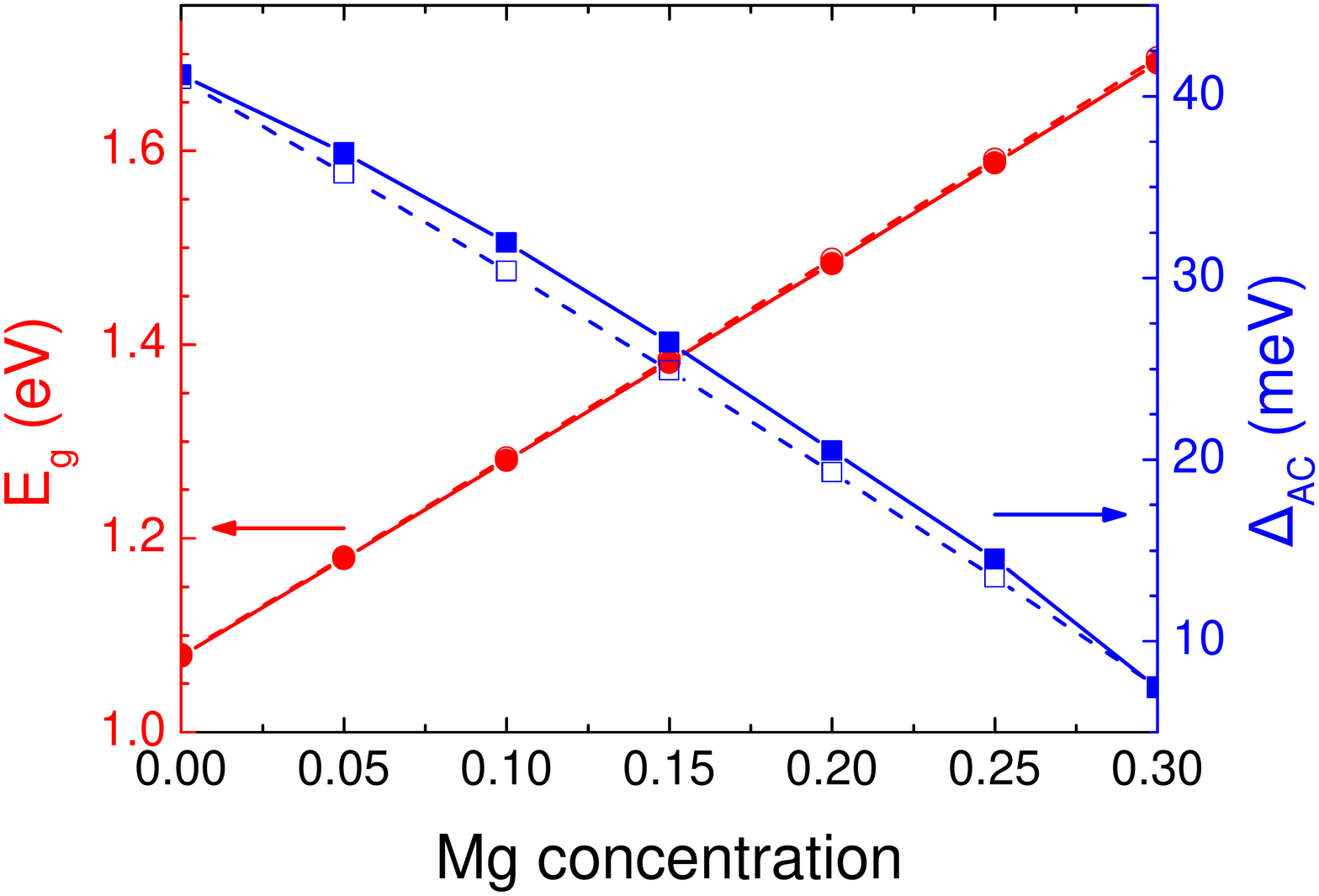}
\caption{\label{img:Es}
Band gap $E_g$ and valence band splitting $\Delta_{AC}$ dependence on the Mg concentration $x$.
(solid lines: fully relaxed, dashed: c-plane grown, the points show the calculated results, while the connecting lines are just a guide for the eye)}
\end{figure}

\begin{figure}
\includegraphics[width=\linewidth]{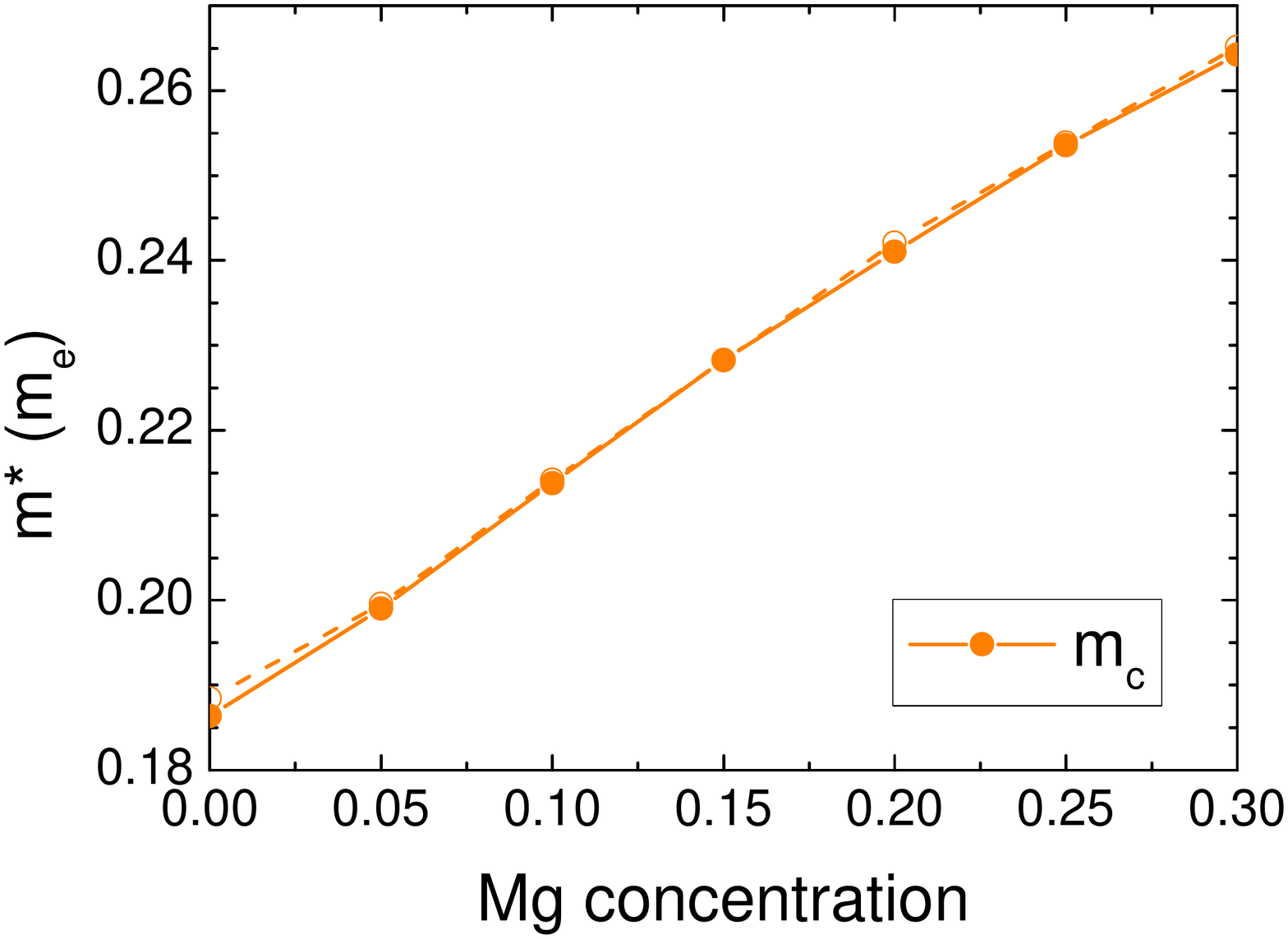}
\caption{\label{img:mc}
Electron effective masses $m_c$ dependence on the Mg concentration $x$.
(solid lines: fully relaxed, dashed: c-plane grown, the points show the calculated results, while the connecting lines are just a guide for the eye)}
\end{figure}

\begin{figure}
\includegraphics[width=\linewidth]{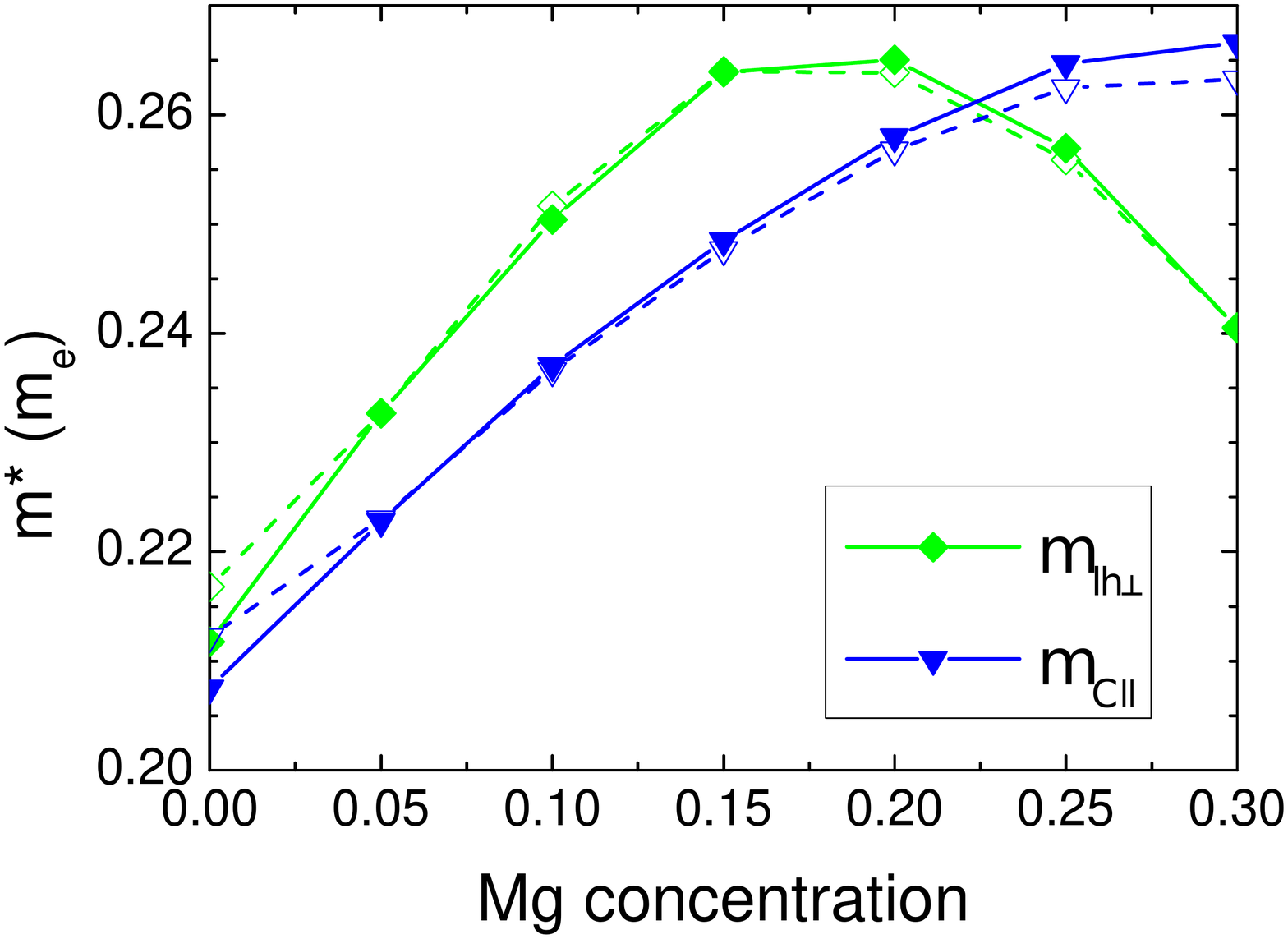}
\caption{\label{img:mlh}
The dependence of the hole effective masses for the light holes on the Mg concentration $x$.
(solid lines: fully relaxed, dashed: c-plane grown, the points show the calculated results, while the connecting lines are just a guide for the eye)}
\end{figure}

\begin{figure}
\includegraphics[width=\linewidth]{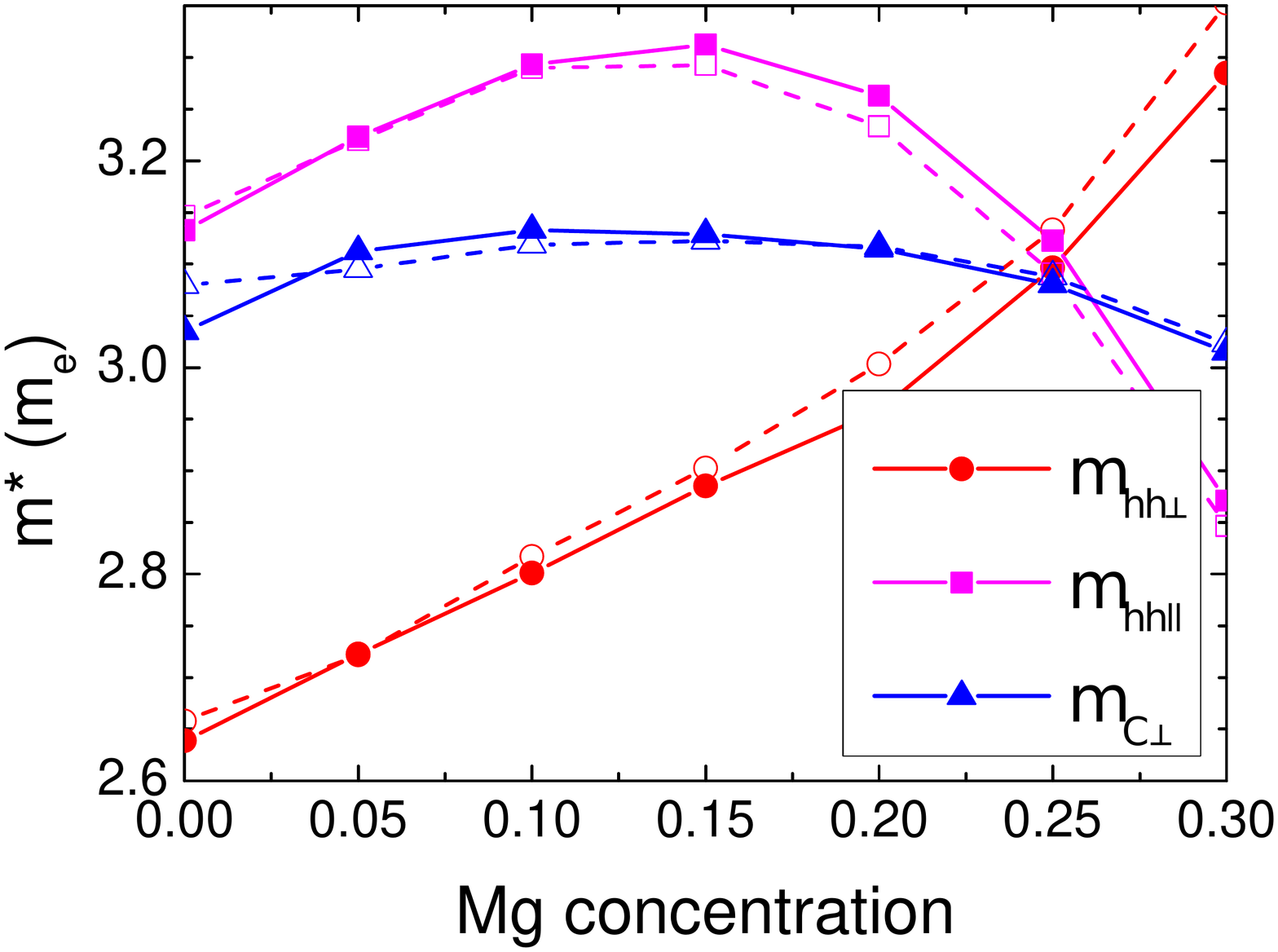}
\caption{\label{img:mhh}
The dependence of the hole effective masses for the heavy holes on the Mg concentration $x$.
(solid lines: fully relaxed, dashed: c-plane grown, the points show the calculated results, while the connecting lines are just a guide for the eye)}
\end{figure}

The results from the fits for pure ZnO are shown in Tab.~\ref{tab:ZnO}. We find that our results for ZnO are similar to other calculated results~\cite{goano2007}. Some of the contributions in the literature include spin-orbit interaction in the calculation. While the electron mass is in rather good agreement with experimental values~\cite{landoltIII}, some of the hole masses show a larger deviation. This is usually found in \textit{ab initio} calculations (see e.g.\ reference~\cite{goano2007,schleife2009,karazhanov2006}) and might be due to the fact that experiments can only measure some direction average of the hole masses~\cite{schleife2009}. The calculated band gap of 1.08 eV is too small, which is common for LDA calculations. More advanced \textit{ab initio} methods are able to predict the band gap more accurately but give similar results for the masses (e.g.\ HSE03+G$_0$W$_0$ in reference~\cite{schleife2009}). On the other hand, the crystal-field splitting is in excellent agreement with experiments~\cite{landoltIII,goano2007}.
We analyze the concentration dependence of the band gap (Fig.~\ref{img:Es}) and find a linear dependence: $E_g/\mathrm{eV}=1.08+2.03\,x$. This linear slope is in outstanding agreement with experimental results. Ohtomo \textit{et al.}\ obtained $E_g/\mathrm{eV}=3.3+2.1\,x$ from transmission spectra for the hexagonal phase and concentrations between 0 and 33\% at room temperature~\cite{ohtomo1998}. Likewise, Chen \textit{et al.}~\cite{chen2003} found $E_g/\mathrm{eV}=3.32+2.00\,x$ at low temperature for concentrations between 0 and 33\% and Wu \textit{et al.}~\cite{wu2010effect} obtained $E_g/\mathrm{eV}=3.384+1.705\,x$ for concentrations between 0 and 44\% at room-temperature from optical absorption spectra. The latter two experiments included the cubic phase at high Mg concentrations and found a separate linear dependence with a different slope for this phase. Our calculated band gap agrees with the LDA result calculated by Maznichenko \textit{et al.}~\cite{maznichenko2009}. The small concentration range considered here does not allow for a sound quadratic fit, but the data supports a small bowing in agreement with the cited experiments.
Further, we find that the crystal-field splitting $\Delta_{AC}$ between the valence bands, shown in Fig.~\ref{img:Es}, decreases approximately linearly over the considered range of concentrations and thus the order of the valence bands will eventually switch. A quadratic extrapolation yields a vanishing crystal-field splitting at ca. $x=0.35$. The concentration dependence of the crystal-field splitting shows a moderate bowing which is notably reduced for the c-plane grown Zn$_{1-x}$Mg$_x$O. The reversal of the order of the valence bands for the pure components (without spin-orbit interaction) was also observed by Schleife \textit{et al.}\ using HSE03+G$_0$W$_0$~\cite{schleife2009}, and by Xu \textit{et al.}\ using the generalized gradient approximation~\cite{xu2008}. Since the spin-orbit interaction introduces an additional splitting of the valence bands, the predicted concentration dependence of the crystal-field splitting cannot be directly observed in experiments. Nevertheless, since the additional spin-orbit splitting is much smaller than the crystal-field splitting, it can be considered a small perturbation. Thus, our result signifies a nonlinear behavior of the valence band splittings. Furthermore, this small perturbation can only have a very weak influence on the effective masses. The splittings are difficult to measure for finite concentrations and this has not yet been verified in experiments~\cite{schmidt2003,teng2000}.
The electron effective mass is presented in Fig.~\ref{img:mc} and we find an increase with the Mg concentration in accordance with experimental results~\cite{cohen2004,lu2006}. A linear fit results in $m_c/m_e=0.186+0.267\,x$. Lu \textit{et al.}~\cite{lu2006} found a linear dependence with a larger slope, while the results by Cohen \textit{et al.}~\cite{cohen2004} suggest a quadratic dependence. All our calculated hole effective masses shown in Fig.~\ref{img:mlh} and Fig.~\ref{img:mhh} have a fair concentration dependence, which is quite different for each band. All hole masses increase for small concentrations $x<0.15$ and some decrease after a maximum in the considered concentration range. A linear interpolation does not seem appropriate for any of the hole masses. We are not aware of any experimental or theoretical investigations of the hole effective masses for finite Mg concentrations. We find a very similar behavior of the effective masses for the two growing conditions considered.

\section{Conclusions}
The influence of the Mg concentration on the band structure and in particular the effective masses of Zn$_{1-x}$Mg$_x$O alloys is analyzed using \textit{ab initio} calculations. We find that the Mg concentration has a strong influence on the band energies and a fair influence on the effective masses. While the band gap increases linearly, the crystal-field splitting decreases. The electron mass increases linearly and the hole masses show a rather complicated and unexpected concentration dependence. Our results are mostly in good agreement with experiments and can be helpful for the simulation of devices.


\begin{acknowledgments}
We thank Prof. B. K. Meyer for helpful discussions. We acknowledge support from the German Science Foundation via grant HE 5922/1-1.
\end{acknowledgments}
\bibliography{bib}
\end{document}